\documentclass[
reprint,
superscriptaddress,
showpacs,
showkeys,
amsmath, amssymb,
aps,
pra,
]{revtex4-1}
\usepackage{booktabs}
\usepackage{mathbbol}
\usepackage{xcolor}
\usepackage{graphicx}
\usepackage{dcolumn}
\usepackage{bm}
\usepackage[colorlinks=true, allcolors=blue]{hyperref}

\usepackage{soul}  

\usepackage{subfigure}
\graphicspath{{figures/}}
\usepackage[separate-uncertainty=false]{siunitx}

\usepackage{xspace}

\newcommand{\bra}[1]{\langle #1\vert}
\newcommand{\ket}[1]{\vert#1\rangle}

\usepackage[normalem]{ulem}

\usepackage{xstring}
\newcommand*{\aref}[1]{%
	\IfBeginWith{#1}{eq:}{Eq.~\eqref{#1}}{}%
	\IfBeginWith{#1}{fig:}{Fig.~\ref{#1}}{}%
	\IfBeginWith{#1}{tab:}{Table~\ref{#1}}{}%
	\IfBeginWith{#1}{appendix:}{Appendix~\ref{#1}}{}%
	\IfBeginWith{#1}{sec:}{Section~\ref{#1}}{}%
}

\begin{document}

	\title{Spin drag and fast response in a quantum mixture of atomic gases }

	\author{Federico Carlini}
	\affiliation{Laboratoire Kastler Brossel, Sorbonne Université, CNRS, ENS-PSL Research University, Collège de France, 4 Place Jussieu, 75005 Paris, France}
	\affiliation{MajuLab, International Joint Research Unit UMI 3654, CNRS, Université Côte d'Azur, Sorbonne Université, National University of Singapore, Nanyang Technological University, Singapore}
	\affiliation{INO-CNR BEC Center and Dipartimento di Fisica, Universit\`a di Trento, 38123 Povo, Italy}
	\author{Sandro Stringari}
	\affiliation{INO-CNR BEC Center and Dipartimento di Fisica, Universit\`a di Trento, 38123 Povo, Italy}
	
	\begin{abstract}
		By applying a sudden perturbation to one of the components of a mixture  of two quantum fluids, we explore the effect on the motion  of the second component on a short time scale. By implementing perturbation theory, we prove that for short times the response of the second component is fixed by the energy weighted moment of the crossed dynamic structure factor (crossed f-sum rule). We also show that by properly monitoring the time duration of the perturbation it is possible to identify  peculiar  fast spin drag regimes, which are sensitive to the interaction effects in the Hamiltonian. Special focus is given to the case of  coherently coupled   Bose-Einstein condensates,   interacting Bose mixtures exhibiting the Andreev-Bashkin effect,  normal Fermi  liquids and the polaron problem. The relevant excitations of the system contributing to the spin drag effect are identified and  the contribution of the low frequency gapless excitations  to the f-sum rule in the density and spin channels is explicitly  calculated employing the proper macroscopic dynamic theories. Both spatially periodic and Galilean boost perturbations are considered.
	\end{abstract}
	
	
	\maketitle
	
	\section {Introduction and general formalism}
	
\begin{figure}
		\begin{center}
			\includegraphics[scale=0.7]{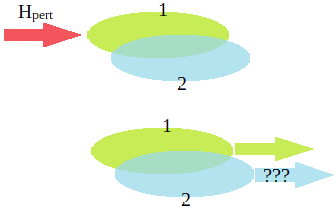}
			\caption{Schematic visualization of fast spin drag. A fast kick, applied to the component 1 of a mixture, can drag the motion of the component 2.}
			\label{fig:fsd}
		\end{center}
	\end{figure}

	Spin drag is an ubiquitous  concept in many branches of physics. It is usually associated with the presence of spin interactions which affect the Euler 
	equation for the spin current. Spin drag can have a collisional nature, giving rise to spin diffusion, or a collisionless nature, causing non dissipative dynamics (see, for example, \cite{Duine,Sommer,Goulko,Fava,Nichols} and references therein). However, spin drag is not necessarily the simple consequence of interaction effects in the equation for the spin current. It can be also caused by the modification of the equation of continuity in the spin channel, due to  the presence of coherent coupling between the two components of the mixture or to beyond mean field effects. This  effect can be explicitly revealed by  applying a sudden perturbation (a kick) to one of the components of the mixture and observing the reaction of the other component on a short time scale (see Fig. \ref{fig:fsd}).   In atomic gases   the perturbation can be experimentally implemented employing  counter propagating laser beams generating selective optical potentials or using suitable position dependent magnetic fields.
	
	Let us suppose that the perturbation applied to the system has the form
	\begin{equation}
	H_{pert}= -\lambda F(1) \Theta(t)
	\label{pert}
	\end{equation}
	where $\Theta(t)$ is the usual Heaviside step function (equal to $0$ for $t<0$ and $1$ for $t\geq 0$) and $F(1)=\sum_{k=1}^{N_1}f({\bf r}_k)$ is a position dependent operator acting on one of the two components of the system (hereafter called component $1$). The question that we  want to address concerns  the short time effects of this perturbation on the component $2$ of the mixture. The shortness time condition is of course related to the frequency of the relevant excitations of the systems and will be explicitly discussed  in the following, together with the conditions that should be physically  imposed to the switching on time of the perturbation (\ref{pert}). 
	
	Using the formalism of linear response theory at zero temperature the changes in the average value $\delta\langle F(2) \rangle=(1/N_2)\int d{\bf r} f({\bf r}) \delta n_2({\bf r},t)$ caused by the perturbation can be written in the form (we set $\hbar=1$):
	\begin{equation}
	\begin{split}
	\delta\langle F(2)\rangle = &{} \frac{\lambda}{N_2}  \sum_n [ \frac{1-e^{-i\omega_{n0}t}}{\omega_{n0}}\bra{0}F(2)\ket{n}\bra{n}F(1)\ket{0}\\ 
	&{}+\frac{1-e^{i\omega_{n0}t}}{\omega_{n0}}\bra{0}F(1)\ket{n}\bra{n}F(2)\ket{0} ] \; ,
	\end{split}
	\label{F2F1}
	\end{equation}
	where  $\omega_{no}$ are the frequencies associated with the states $\ket{n}$ excited by the perturbation (\ref{pert}). For simplicity, we have assumed here that the operator $F$ is Hermitian. Of course the changes  in the value of $\langle F(1)\rangle$ are simply obtained by replacing the operator $F(2)$ with $F(1)$ and $N_2$ with $N_1$ in the above equation. Let us notice that Eq. (\ref{F2F1}) is valid in general for any   pair of Hermitian operators. Useful choices for  $f({\bf r})$ that will be discussed in the paper  are  $f({\bf r})= \cos(qx)$ and $f({\bf r}) =x$. The first choice is relevant in the case of perturbations  produced by applying two counter propagating  laser fields,  $q$ being the typical momentum transferred to the system. The second one is instead particularly relevant for harmonically trapped configurations, where the perturbation is provided by the  sudden shift of the center of the trapping potential. Alternatively it can be provided by a fast spin selective boost generated by an optical potential.  
	
	A first immediate result emerging from Eq.(\ref{F2F1}) is that, if  all the excitation frequencies $\omega_{n0}$ relative to the states 
 contributing to the relevant matrix elements satisfy the condition $\omega_{n0}t \ll 1$, then the expansion of $\exp{(\pm i\omega_{n0}t)}$ yields the following expression for the crossed response (hereafter called ultrafast spin drag regime)
	\begin{equation}
	\begin{split}
	\delta\langle F(2)\rangle^{ ultra} = &{} \frac{\lambda t^2}{N_2}\int d\omega \omega S_{1,2}(\omega)  \\ 
	=&{}  \frac{\lambda t^2}{2 N_2} \langle [F(2),[H,F(1)]]\rangle \; ,
	\end{split}
	\label{FHF}
	\end{equation}
	where
	\begin{equation}
	\begin{split}
	S_{1,2}(\omega) =&{} \frac{1}{2}\sum_n\delta(\omega - \omega_{no})\times(\bra{0}F(1)\ket{n}\bra{n}F(2)\ket{0}\\
	&{}+\bra{0}F(2)\ket{n}\bra{n}F(1)\ket{0})
	\label{S12}
	\end{split}
	\end{equation}
	is the crossed dynamic structure factor relative to the operators $F(1)$ and $F(2)$ and
	we have used the completeness relation allowing for an easy calculation of the  response in terms of  commutators. In the same ultrafast regime the response experienced by the first component is instead  given by
	\begin{equation}
	\delta\langle F(1)\rangle^{ultra} = \frac{\lambda t^2}{2 N_1} \langle [F(1),[H,F(1)]]\rangle \; .
	\label{F1HF1}
	\end{equation}
	The expansion (\ref{FHF}) shows that,  if the crossed double commutator   does not vanish,  the perturbation, applied to the first component, causes the emergence of a dynamic flow $ d\langle F(2)\rangle/dt$ in the second one (see  Fig.\ref{fig:fsd}). The above results share interesting analogies with the large $\omega$ behavior of the usual dynamic polarizability, providing the response of the system to a perturbation switched on adiabatically and varying in time like $e^{\eta t}e^{-i\omega t}$ with $\eta \to 0^+$ \cite{Book}, the inverse of the waiting time $t$ replacing  the   role of  the  frequency $\omega$ characterizing the usual dynamic response.
	
The double commutators  of Eqs. (\ref{FHF}) and (\ref{F1HF1})  are easily evaluated once the Hamiltonian of the system is known.  For momentum  independent interactions and in the absence of coherent coupling between the two components, the crossed double commutator $[F(2), [H,F(1)]]$ identically vanishes, as the only term contributing to the commutator $[H,F(1)]$ is the kinetic energy of the first component. As a consequence the spin drag effect is absent in this case. 
Even in the absence of the ultrafast spin drag effect (\ref{FHF}), there exist interesting spin drag effects taking place at longer times, as we will discuss in the following. Indeed, in the presence of a significantly large frequency gap in the excitation spectrum, one can separate, in the sum (\ref{F2F1}), a low energy contribution, arising from the states with excitation energies below the frequency gap (in the following indicated as $\omega_{low}$),  from a high energy contribution arising from the states  above the frequency gap (in the following  indicated as $\omega_{high}$). For time durations of the perturbation satisfying the condition $2\pi/\omega_{high}\ll t\ll 2\pi/\omega_{low}$ (hereafter called fast spin drag  regime),  the response (\ref{F2F1}) reduces to  the sum of two contributions:
	\begin{equation}
	\delta\langle F(2)\rangle^{fast}=\delta\langle F(2)\rangle_{low} + \delta\langle F(2)\rangle_{high} \; .
	\label{lh}
	\end{equation}
	In Eq. (\ref{lh}) we   identify  the term
	\begin{equation}
	\delta\langle F(2)\rangle_{low} =\frac{\lambda t^2}{N_2} m_1(1,2)_{low} \;  ,
	\label{m1low}
	\end{equation}
	where $m_1(1, 2)_{low}$ is the contribution given by the low energy excitations to the energy weighted moment 
	\begin{equation}
	m_1(1,2) = \int d\omega \omega S_{1,2}(\omega)
	\label{m1S12}
	\end{equation}
	of the crossed dynamic structure factor (\ref{S12}). 	
	The term $\delta\langle F(2)\rangle_{high}$ can be instead easily calculated taking the time average of the signal over times much longer than the inverse of the high energy excitation frequencies and, of course, much smaller than the inverse of the low energy ones. In this case, the oscillating terms $\exp{(\pm i\omega_{n0}t)}$ of these high energy states can be ignored and the second term in Eq. (\ref{lh}) takes the simplified form
	\begin{equation}
	\delta\langle F(2)\rangle_{high} = \frac{2\lambda}{N_2}   m_{-1}(1,2)_{high} \; ,
	\label{m-1highst}
	\end{equation}
where $m_{-1}(1,2)_{high}$ is the contribution to the inverse energy  weighted moment of the dynamic structure factor (\ref{S12}) (static crossed polarizability) arising from the high energy states. Notice that result  (\ref{m-1highst})  is a static effect which does not produce any dynamic flow $d \langle F(2)\rangle/d t$. Result (\ref{m-1highst}) can be equivalently  obtained by assuming that the perturbation (\ref{pert}) is not switched on instantaneously at $t=0$, but within a transient ramping time $\tau$ which should be  larger (smaller) than the inverse of the frequencies of the high (low) energy states. If  the high energy states form a continuum, result (\ref{m-1highst}) can be also the consequence of a damping effect occurring for times longer than the inverse of the corresponding width. This is the case, for example, of beyond mean field multi-pair excitations  in Fermi liquids  or in superfluid mixtures.

\begin{figure}
		\begin{center}
			\includegraphics[scale=0.38]{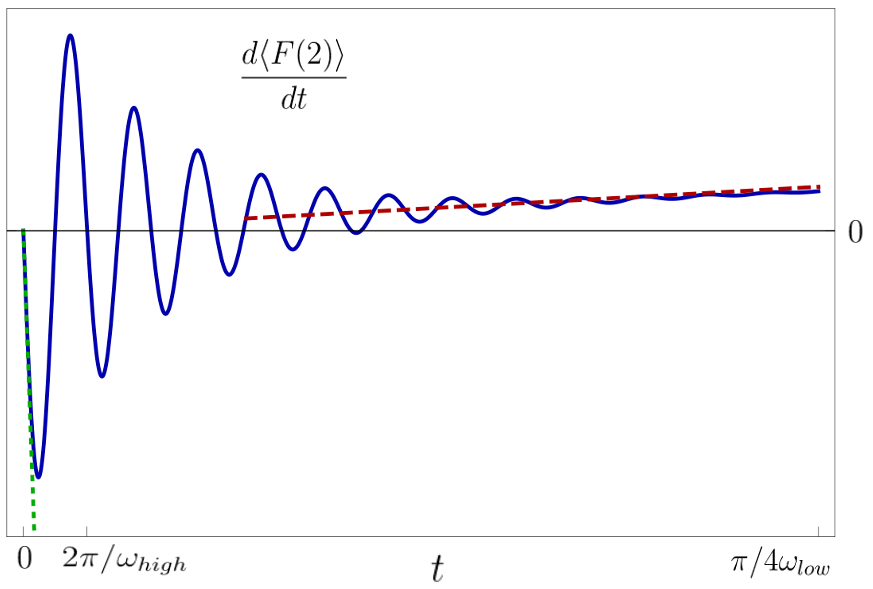}
			\caption{Schematic representation (blue line) of the time evolution of the physical signal $d \langle F(2)\rangle/d t$ caused by the perturbation (\ref{pert}) applied to the component 1. The red dashed line highlights the fast spin drag regime response. The oscillations with frequency $2\pi/\omega_{high}$ are damped due to the presence of a finite ramping time. The slope at $t\ll 2\pi/\omega_{high}$ highlights the ultrafast spin drag effect (green dotted line). In the case of the figure the slope is negative, corresponding to the coherently coupled configuration discussed in Sec. \ref{cc}.}
			\label{fig:fast}
		\end{center}
	\end{figure}

In conclusion, we have identified two different regimes (ultrafast and fast, respectively) revealing the possible existence of peculiar spin drag effects. In Fig.\ref{fig:fast} we schematically represent an example of time evolution of the physical signal $d \langle F(2)\rangle/d t$ caused by the perturbation (\ref{pert}) applied to the component 1. This corresponds to the dynamic flow whose measurement could explicitly provide the spin drag effect. For times of the order of $2\pi/\omega_{high}$ the signal is characterized by oscillations with frequency $\omega_{high}$, whose amplitude is suppressed because of the effects discussed above. The {\it ultrafast}  effect (\ref{FHF}), when present, takes place for $t\ll 2\pi/\omega_{high}$. In the case of Fig.\ref{fig:fast} it is characterized by a negative slope.  The {\it fast} spin drag effect, taking place for $2\pi/\omega_{high}\ll t\ll2\pi/\omega_{low}$,  is instead revealed  by the fact that the  oscillations of the signal $d \langle F(2)\rangle/d t$, if not already damped out, take place  around an average value increasing linearly in time, in accordance with the quadratic time dependence of Eq. (\ref{lh}). In Fig.\ref{fig:fast} the fast spin drag is represented by the red dashed line. The signal will start oscillating at larger times, of the order of $2\pi/\omega_{low}$. Since the fast spin drag effect involves only the excitation of the low frequency part of the spectrum, it can be explicitly calculated employing the proper macroscopic dynamic theory e.g. the hydrodynamic formalism for superfluid mixtures or the Landau's theory for interacting Fermi mixtures in the normal phase.

The distinction between the high and low frequency excitations, characterizing the dynamic structure factor and the spin drag response discussed above, depends on the nature of the system considered. In some cases (see for example quantum Bose mixtures without coherent coupling  and normal Fermi systems) the relevant high frequency states, excited by operators of macroscopic nature such as, for example, operators of the form $\cos(qx)$ with small $q$, or of the dipole form $x$,  are the results of beyond mean field effects, which are responsible for a non-vanishing  contribution to the f-sum rule in the spin channel. In the case of Bose mixtures with coherent coupling these high energy states are instead naturally caused by the gap produced by the coupling (for example  Raman coupling in spin-orbit Bose-Einstein condensed gases). The low frequency excitations considered in this work  have in general a gapless nature. In the case of Bose superfluids they corresponds to usual hydrodynamic phonons, while in  Fermi liquids they correspond to zero sound and/or to particle-hole excitations. In the presence of harmonic trapping the  frequencies of the low energy states excited by the dipole operator  are instead of the order of the trapping frequency. In ultracold atomic gases the separation between the high and low  frequency excitations discussed above corresponds in most cases to  realistic  experimental conditions. 
	
    Let us finally notice that, if one studies a balanced mixture ($N_1=N_2=N/2$) characterized by an Hamiltonian symmetric in the variables 1 and 2 of the two components, the responses $\delta\langle F(1)\rangle$ and $\delta\langle F(2)\rangle$  caused by the   perturbation (\ref{pert}) can be conveniently written in terms of the density and the spin density responses to sudden perturbations of the form $ - \lambda F_{n(\sigma)}  \Theta(t)$, where $F_n$ and $F_{\sigma}$ are, respectively, the in phase (density) and out of phase (spin density) combinations of $F(1)$ and $F(2)$. One can indeed write:
    \begin{equation}
	\begin{split}
	\delta\langle F(1)\rangle= \frac{1}{2}(\delta\langle F_n\rangle +\delta\langle F_{\sigma}\rangle)\\
	\delta\langle F(2)\rangle= \frac{1}{2}(\delta\langle F_n\rangle -\delta\langle F_{\sigma}\rangle)
	\end{split}
	\label{FnFs} 
	\end{equation}
	where the quantities $\delta\langle F_n\rangle$ and $\delta\langle F_{\sigma}\rangle$ are normalized to the total number $N$ of particles.\\ 
	As one can see from the above equation, the balance between the density and spin density responses  fixes also the sign of the dynamic flow   experienced by the second component with respect to the flow of the first component to which the sudden   perturbation was initially applied.
	
	It is also useful to consider the contributions arising from the low and high frequency excitations  of the system to the relevant energy weighted moment 
	\begin{equation}
	\begin{split}
	m_1(F_{n(\sigma)}) &{}=\int d\omega S_{n(\sigma)}(\omega) \omega\\
	&{}=\sum_n|\langle 0|F_{n(\sigma)}||n\rangle|^2 \omega_{n0}
	\end{split}
	\label{m1ns}
	\end{equation}
	of the in phase and out of phase  dynamic structure factor. 
	As discussed in the first part of this section and in view of Eq. (\ref{FnFs}),  these contributions  are   relevant for the calculation of the spin drag effect $\delta\langle F(2)\rangle$. Their behavior is  schematically reported in the Tables \ref{twithcc} and \ref{twithoutcc} in the case of a momentum dependent perturbation.
	
	\section{ Coherently coupled Bose-Einstein condensed gases}
	\label{cc}
	
	 Coherently coupled Bose-Einstein condensed gases represent an interesting case where the spin drag regimes discussed in the introduction can be explicitly investigated. 
	
	In the case of spin-orbit coupled configurations the value of the high energy gapped excitations is fixed by the experimentally controllable Raman coupling $\Omega$ entering the   single particle Hamiltonian
	\begin{equation}
	H_{sp}^{SOC}=\frac{1}{2m}\big[(p_x- k_0\sigma_z)^2+p^2_\perp\big]-\frac{\Omega}{2}\sigma_x +V_{ext}
	\label{HSOC}
	\end{equation}
	while the spin-orbit term, fixed by the momentum transfer $k_0$ between the two counter propagating laser fields generating the coherent coupling, is at the origin of non trivial effects affecting  the gapless branch of the system and hence the velocity of sound. The value of $k_0$ is practically vanishing in the case of radio frequency or microwave coupling, hereafter called Rabi coupling, where it can be safely set equal to zero.
	
	The energy weighted moments relative to the density $F_n=F(1)+F(2)$ and spin density operators $F_{\sigma}=F(1)-F(2)$ are easily evaluated and take, respectively,  the form
	\begin{equation}
	m_1(F_n) = \frac{1}{2}\langle [F_n,[H,F_n]]\rangle=\frac{1}{2 m} \int d{\bf r} |\nabla f({\bf r})|^2n({\bf r}) 
	\label{m1SOCn}
	\end{equation}
	and 
	\begin{equation}
	\begin{split}
	m_1(F_\sigma) =&{}\frac{1}{2} \langle [F_\sigma,[H,F_\sigma]]\rangle= \Omega \int d{\bf r} f^2({\bf r})n({\bf r})\\
	&{}+\frac{1}{2 m}\int d{\bf r} |\nabla f({\bf r})|^2n({\bf r}) \; ,
	\end{split} 
	\label{m1SOCsigma}
	\end{equation}
	showing that, while   in the density channel the energy weighted sum rule is not affected by the Raman coupling, in the spin channel the coupling provides a crucial contribution which, according to Eq.(\ref{FnFs}), is at the origin of a spin drag effect in the ultrafast regime.  By considering, for simplicity, the case of the spatially  periodic perturbation $f(\mathbf{r})=\cos(qx)$, one finds the result 
	\begin{equation}
	\frac{d \langle F(2)\rangle}{d t}^{ultra}= -\frac{\lambda t}{2} \Omega \; ,
	\label{FHFbis}
	\end{equation}
	while, for the first component, one finds
	\begin{equation}
	\frac{d \langle F(1)\rangle}{d t}^{ultra}=\frac{\lambda t}{2}\bigg(\frac{q^2}{m}+\Omega\bigg)
	\label{FHFbis1}
	\end{equation}
	where we used $\langle \cos^2(q x)\rangle =1/2$. Notice that, interestingly, the two components will be put in motion in opposite direction.
	
	The fast regime, taking place for times satisfying the condition $t\Omega>2\pi$, but shorter than the inverse of the phonon frequencies,  instead involves the dynamic excitation  of only the phonon modes, which are properly described  by   hydrodynamic (HD) theory. In the case of spin-orbit coupled BECs the hydrodynamic formalism was developed in \cite{martone} (see also \cite{SOCreview}) and exhibits very peculiar features.   In the plane wave and single minimum phases the excitations with small frequencies are indeed characterized by the locking of the relative phase   between the two  components. In the linear limit of small amplitude oscillations of macroscopic nature, where one can safely ignore quantum pressure effects, the phase locking  reduces the coupled Gross-Pitaevskii equations to the following set of hydrodynamic equations:  
	
	\begin{equation}
	\partial_t \delta n + \frac{1}{m}{\bf \nabla} \cdot ( n {\bf \nabla} \Phi)-\frac{k_0}{m}\nabla_x \delta s_z=0 
	\label{HD1}
	\end{equation}
	
	\begin{equation}
	\partial_t\nabla \Phi+ \nabla (g \delta n) =0
	\label{HD2}
	\end{equation}
	and
	\begin{equation}
	-\frac{k_0}{m}n\nabla_x\Phi+  \frac{\Omega}{2}    \delta s_z =0
	\label{HD3}
	\end{equation} 
	where $\delta n$ and $\delta s_z$ are the fluctuations in the total density and in the spin density taking place during the oscillation.
	In deriving the above equations we have assumed, for sake of simplicity, that the values of  the intra-species and inter-species interaction coupling constants coincide and are given by the parameter $g$.  Furthermore we have restricted  the derivation to the single minimum phase taking place for values of the Raman coupling satisfying the condition $\Omega \ge \Omega_c\equiv 2k^2_0/m$ and characterized, at equilibrium, by the vanishing of the phase $\Phi$ of the order parameter and of the spin density $s_z$. 
	\begin{table*}[t]
		\caption{Mixtures with coherent coupling. Momentum-$q$ dependence of strength, excitation energy and contribution to the energy weighted sum rule $m_1$, for excitation operators of the form $F(q)=\sum_k\cos(q x_k)$. The q-dependence is given for low and high frequency excitations in both the density and spin channels for unpolarized   spin-orbit (a) and  Rabi (b) coupled configurations.  In the Rabi case, due to the absence of hybridization between the density and spin degrees of freedom, the low and high energy excitations have a distinct density and spin nature, respectively.\\[-6ex]}
		\label{twithcc}
	\end{table*}
	
	\begin{table}[t]
		\setcounter{table}{0}
		\renewcommand{\thetable}{\Roman{table}a}
		\caption{SOC coupling}
		\label{twithcca}
		\begin{tabular}{c c c c c c c c}
			\hline
			\\[-1ex]
			& \multicolumn{2}{c}{{\bf density}}& & & &\multicolumn{2}{c}{{\bf spin}}\\
			& \multicolumn{1}{c}{low} & \multicolumn{1}{c}{high} & & & & \multicolumn{1}{c}{low} & \multicolumn{1}{c}{high}\\[+1ex]
			\cline{2-3}\cline{7-8}
			\\[-1ex]
			\boldmath$|\langle 0|F(q)|n\rangle|^2$\unboldmath & $q$ & $q^2$ & & & & $q$ & const \\[+1ex]
			\boldmath$\omega_{n0}$\unboldmath & $q$ & $\Omega$ & & & & $q$ & $\Omega$ \\[+1ex]
			\boldmath$m_1$\unboldmath& $q^2$ & $q^2$ & & & & $q^2$ & const\\[+1ex]
			\hline
		\end{tabular}
	\end{table}
	The first equation is the equation of continuity, which is deeply affected by spin-orbit coupling, reflecting the fact that the physical current is not simply given by the gradient of the phase as happens in usual superfluids, but contains a crucial spin dependent term, which implies that  even in the density channel the f-sum rule is not exhausted by the gapless phonon excitation (see Table \ref{twithcca}). The second equation corresponds to the Euler equation and fixes the time dependence of the phase gradient  of the order parameter. Finally, the third equation follows from the variation of the action with respect to the spin density and is responsible for the hybridization between the density and spin density degrees of freedom in the propagation of sound \cite{martone}. This equation characterizes in a unique way the consequences of spin-orbit coupling. It identically vanishes in the case of radio frequency or microwave coupling where the phonon modes are pure density waves.
	
	The linearized equations of motion (\ref{HD1}-\ref{HD3}) can be rewritten in the useful form:
	\begin{equation}
	\partial_t^2 \delta n = \frac{g}{m}[\nabla_\perp\cdot(n \nabla_\perp \delta n)+\frac{m}{m^*}\nabla_x(n \nabla_x \delta n)]
	\label{linHD}
	\end{equation}
	where $n$ is the equilibrium density and  we have introduced the effective mass  $m/m^*=1-\Omega_c/\Omega$. One can show that Eq.  (\ref{linHD}) holds also in the plane wave phase, taking place for $\Omega < \Omega_{c}$. In this case the effective mass is given by $m/m^*=1-(\Omega/\Omega_c)^2$. 
	In uniform matter ($V_{ext}=0$) Eq.(\ref{linHD}) provides, along x, the phonon  dispersion law  $\omega=cq$ with $c^2=g n/m^*$, revealing a strong reduction of the sound velocity in the vicinity of the second order phase transition between the plane wave and the single minimum phase, where the effective mass has a divergent behavior.
	
	By applying a perturbation of the form $-\lambda F_n \Theta(t) $ (density perturbation), the hydrodynamic equation (\ref{HD2})  contains the additional term $-\lambda \nabla f({\bf r})\Theta(t)$, while in the case of the spin perturbation $-\lambda F_\sigma \Theta(t)$ the  term $-\lambda f({\bf r})\Theta(t)$ should be added to the third  hydrodynamic equation (\ref{HD3}). We first consider  the case of uniform matter and we make the choice $f({\bf r})=\cos (q x)$ for the position dependence of the excitation operator. The solution of the hydrodynamic equations then yields the following result for the density  
	\begin{equation}
	\delta\langle F_n \rangle^{HD}
	=\frac{\lambda}{2 g n}\bigg[1-\cos(cqt)\bigg]
	\label{HDn}
	\end{equation}
	and spin density  
	\begin{equation}
	\delta\langle F_{\sigma}\rangle^{HD}
	=\frac{\lambda}{\Omega}\frac{m^*}{m}\bigg[1-\frac{\Omega_c}{\Omega}\cos(cqt)\bigg]
	\label{HDs}
	\end{equation}
	linear responses to the  step function time dependent perturbation. 
	Some comments are in order here:
	
	- By taking the time average $\overline{\cos (cqt)}=0$,  one recovers the static response of the system, given by the compressibility $1/gn$ in the density case, and by the magnetic susceptibility $(m^*/m)(2/\Omega)$ in the spin case.
	
	- By expanding the cosine for times shorter than the inverse of the phonon frequency and taking the difference (\ref{FnFs}) between the density and spin density responses (\ref{HDn}-\ref{HDs})  
	one finds that  the dynamic flow  of the  second component, caused by a fast perturbation applied to the first one,  takes the form  
	\begin{equation}
	\frac{d \langle F_2\rangle}{d t}^{fast}
	=\frac{\lambda t}{4} \frac{q^2}{m}\bigg[\frac{m}{m^*}-\frac{2 g n\Omega_c}{\Omega^2}\bigg]\; .
	\label{HDdrag}
	\end{equation}
	
	Vice versa, the dynamic flow of the first component is given by
	\begin{equation}
	\frac{d  \langle F_1\rangle}{d t}^{fast}
	=\frac{\lambda t}{4} \frac{q^2}{m}\bigg[\frac{m}{m^*}+\frac{2 g n\Omega_c}{\Omega^2}\bigg]\; .
	\label{HDdrag1}
	\end{equation}
\begin{table}[t]
    \setcounter{table}{0}
	\renewcommand{\thetable}{\Roman{table}b}
	\caption{Rabi coupling}
	\label{twithccb}
	\begin{tabular}{c c c c c c }
		\hline\\[-1ex]
		& \multicolumn{1}{c}{{\bf density}}& & & &\multicolumn{1}{c}{{\bf spin}}\\
		& \multicolumn{1}{c}{low}  & & & & \multicolumn{1}{c}{high}\\[+1ex]
		\cline{2-2}\cline{6-6}\\[-1ex]
		\boldmath$|\langle 0|F(q)|n\rangle|^2$\unboldmath & $q$ & & & &  const \\[+1ex]
		\boldmath$\omega_{n0}$\unboldmath & $q$ & & & & $\Omega$ \\[+1ex]
		\boldmath$m_1$\unboldmath& $q^2$  & & & & const\\[+1ex]
		\hline
	\end{tabular}
\end{table}

	Note that, although in this section we have mainly focused on the single minimum phase, the formalism can be easily extended  to the plane wave phase of the spin-orbit coupled configuration. It is also worth noticing that the above results can be immediately applied to the case of radio-frequency  coupled mixtures, where the momentum transfer is negligible, by setting $k_0=0$, and hence $m^*=m$ and $\Omega_c=0$. In the case of Rabi coupling, the behavior of the drag effect in the fast regime reflects the fact that  the f-sum rule in the density channel is exhausted by the phonon mode  (see Table \ref{twithccb}) and has opposite sign compared to  result (\ref{FHFbis}) holding in the ultrafast regime, where the energy weighted sum rule in the spin channel is dominated, at small $q$, by the high frequency gapped state. In the   spin-orbit Raman coupled case, the dynamic excitation of the spin degree of freedom (second term in rhs of Eq.(\ref{HDs})) can  easily become the dominant effect also in the fast regime, if  $\Omega$ is close enough to the transition to the plane wave phase, where $m^*/m$ takes a large value. Indeed, in this case the low frequency contribution to the energy weighted sum rule in the spin channel (behaving like $q^2$,  see Table \ref{twithcca}) becomes dominant as compared to the corresponding contribution in the density channel.
	
	Similar results are obtained if one considers a perturbation of the form $f({\bf r})=x$ applied to the component 1 and providing, for short times, a Galilean boost. In this case the time derivatives of $\langle F(1)\rangle$ and $\langle F(2)\rangle$ coincide with the center of mass velocities $V_1$ and $V_2$ of the  corresponding atomic clouds  and the spin drag effect is well characterized by the ratio
\begin{equation}
        \bigg( \frac{V_2}{V_1}\bigg)^{ultra}= -\frac{\Omega\langle x^2\rangle}{\Omega\langle x^2\rangle + 1/m}
\label{ULTRASOC}
\end{equation}
in the ultrafast regime and 
	\begin{equation}
    \bigg( \frac{V_2}{V_1}\bigg)^{fast}= \frac{\langle m/m^*-2gn\Omega_c/\Omega^2\rangle}{\langle m/m^*+2gn\Omega_c/\Omega^2\rangle}
\label{FSOC}
\end{equation}
in the fast regime. In the case of Rabi coupling, where $\Omega_c=0$ and $m^*=m$, Eq. (\ref{FSOC}) reduces to the value $(V_2/V_1)^{fast}=1$ having opposite sign compared to the result holding in the ultrafast regime (\ref{ULTRASOC}).

In the presence of harmonic trapping the averages appearing in Eq.s (\ref{ULTRASOC}) and (\ref{FSOC}) take into account the spatial inhomogeneity of the density profiles.

\section {Andreev-Bashkin effect}
	
The Andreev-Bashkin (AB) effect \cite{AB} is a beyond mean field phenomenon, where collisionless spin drag takes place because of the presence of current dependent interactions between the two superfluids forming the quantum mixture. These interactions modify the equation of continuity in the spin channel and, in the case of two interacting Bose-Einstein condensates, are not accounted for by mean field  Gross-Pitaevskii theory. They can be properly  included in the hydrodynamic theory of superfluids in order to correctly describe  the low frequency macroscopic dynamics of the mixture at  zero temperature.
	
	Due to Galilean invariance the equation of continuity for the total density is not modified by the inclusion of beyond mean field effects and takes the usual form  
	\begin{equation}
	\partial_t \delta n + \frac{1}{m}{\bf \nabla} \cdot ( n {\bf \nabla} \Phi)=0 
	\label{ABHD1}
	\end{equation}
	where $n=n_1+n_2$ is the total density and $\Phi=(\phi_1+\phi_2)/2$ is the global phase of the order parameter. The equation of continuity for the spin channel is instead affected by the interactions between the two components of the mixture being characterized by the presence of a novel spin drag term fixed by the spin drag density $n_{drag}$
	\begin{equation}
	\partial_t \delta s + \frac{1}{m}{\bf \nabla} \cdot ( n {\bf \nabla} \Phi_s)-\frac{4}{m}{\bf \nabla} \cdot ( n_{drag}{\bf \nabla} \Phi_s)=0 
	\label{ABHD3}
	\end{equation}
	where $\Phi_s=(\phi_1-\phi_2)/2$ is the relative phase of the two order parameters. In writing  the above equations we have assumed, for simplicity, $m_1=m_2 \equiv m$ and $n_1=n_2=n/2$ at equilibrium. As discussed in \cite{recati} the spin drag term  $n_{drag}$ is directly connected with the occurrence of  novel non diagonal terms in the superfluid density. 
	
	The hydrodynamic equations for the gradient of the total and relative phases are governed, respectively,  by the compressibility and by the magnetic susceptibility of the mixture 
	
	\begin{equation}
	\partial_t\nabla \Phi+ \nabla \frac{\delta n}{2\kappa_d} =0 \; .
	\label{ABHD2}
	\end{equation}
	
	and
	\begin{equation}
	\partial_t\nabla \delta \Phi+ \nabla \frac{\delta s}{2\chi_s} =0
	\label{ABHD4}
	\end{equation}
	In  the mean field approach applied to a weakly interacting Bose-Bose mixture, one can simply write $\kappa_d=1/(g+g_{12})$ and $\chi_s=1/(g-g_{12})$ where $g$ and $g_{12}$ are, respectively, the intra-species and inter-species interaction coupling constants entering the Gross-Pitaevskii equation for the mixture \cite{Book}. In the same mean field scheme one has $n_{drag}=0$ so that, in order to make the whole hydrodynamic picture consistent, one should include beyond mean field effects also in the calculation of the compressibilty and of the magnetic  susceptibility \cite{Romito20}. 
	
	The equation of continuity (\ref{ABHD3}) for the spin density directly reflects the occurrence of the spin drag effect and is responsible for a new behavior of the energy weighted sum rule.  While in the density channel ($F_n= F(1)+F(2)$)  this sum rule takes the usual value 
	\begin{equation}
	m_1^{HD}(F_n) =\frac{1}{2 m} \int d{\bf r} |\nabla f({\bf r})|^2n({\bf r})  \; ,
	\label{m1ABn}
	\end{equation}
	in the spin channel ($F_\sigma=F(1)-F(2)$ ) one finds the result
	\begin{equation}
	\begin{split}
	m_1^{HD}(F_\sigma) = &{}
	\frac{1}{2 m}\int d{\bf r} |\nabla f({\bf r})|^2n({\bf r})\\ &{}-\frac{2}{m}\int d{\bf r}  |\nabla f({\bf r})|^2 n_{drag}({\bf r})
	\label{m1ABsigma}
	\end{split}
	\end{equation}
which  reflects the fact that the low energy excitations do not exhaust the  energy weighted sum rule in the spin channel. By the way, result (\ref{m1ABsigma}) imposes the condition $n_{drag} < n/4$ which should be satisfied in order to ensure the positiveness of the $m_1^{HD}$ sum rule. At the same time, since $m_1^{HD}$ cannot exceed the total f-sum rule one should have $n_{drag} >0$. When evaluated with the proper microscopic many-body Hamiltionian, which yields a vanishing value for the crossed  double commutator $[F(2),[H,F(1)]]$, the spin sum rule $m_1(F_\sigma)$   would actually coincide with  
	the sum rule result (\ref{m1ABn}) holding in general in the density channel, beyond the hydrodynamic description. 
	This behavior reveals  the existence, at higher energies, of excitations which provide the remaining difference to the energy-weighted sum rule in the spin channel. This effect, here derived in superfluid Bose mixtures exhibiting the AB effect,  was first pointed out in the framework of the normal phase of liquid $^3$He \cite{PhysRevLett.37.845} (see also next section).
	
	In a weakly interacting Bose gas mixture the energy of these states is expected to be of the order of the chemical potential, but a quantitative  estimate, based on a microscopic calculation, is not so far available.
	
	In the both AB effect and in coherently coupled BECs the occurrence of high energy excitations plays an important role in the behavior of the hydrodynamic contribution to the energy weighted sum rules. However, important differences emerge in the two cases:
	
	- While in the SOC case the physical origin of the  high energy gapped states is due to the Raman coupling $\Omega$ caused by the external laser fields and has consequently a single-particle nature, in the AB effect the existence of these high energy states is the consequence of beyond mean field  many-body effects.
	
	- A second important difference is that in the SOC case the violation of the f-sum  rule result $(1/2m) \int d{\bf r} |\nabla f({\bf r})|^2n({\bf r})$ in the hydrodynamic description does not concern only the spin, but also the density channel and results in a huge effect near the transition between the plane wave and the single minimum phase. The AB effect instead concerns only the spin channel, the f-sum rule being exactly exhausted in the density channel by the low frequency hydrodynamic mode, as a consequence of Galilean invariance.
	
	- Contrary to the SOC case, in the AB effect there is no spin drag in the ultrafast regime. Indeed, as mentioned above,  the crossed  double commutator $[F(2),[H,F(1)]]$  identically vanishes, when evaluated employing  the microscopic Hamiltonian.  
	
	- Another important difference concerns  the structure of the hydrodynamic theory in the two cases: in the SOC case the locking of the relative phase of the two condensates causes the existence of a single gapless excitation branch and the density and the spin degrees of freedom are strongly hybridized, loosing their independent nature. In the AB effect the density and spin degrees of freedom are instead fully decoupled and the system exhibits two gapless excitation branches. 
	
	- A final difference concerns the fact that the high energy states in the AB effect do not contribute to the static magnetic susceptibility, differently from what happens in the SOC case. 
	
	If we switch on a perturbation of the form $-\lambda F\Theta(t)$, with a proper choice of the ramping time, which should be larger than   the inverse of the high energy frequency excitations, then the hydrodynamic theory outlined above is well suited to describe the response of the system. In uniform matter, making the choice $f({\bf r})=\cos(qx)$, the hydrodynamic approach yields the result 
	
	\begin{equation}
	\delta\langle F_n \rangle^{HD}
	=\frac{\lambda\kappa_d}{n}\bigg[1-\cos(c_d q t)\bigg]
	\label{ABHDn}
	\end{equation}
	for the response in the density channel ($F=F_n$). In the spin channel ($F=F_\sigma$) the response is instead given by 
	\begin{equation}
	\delta\langle F_{\sigma}\rangle^{HD} 
	=\frac{\lambda \chi_s}{n}\bigg[1-\cos(c_s q t)\bigg]
	\label{ABHDs}
	\end{equation}
	where $c_d$ and $c_s$ are, respectively, the density and spin sound velocities 
	\begin{equation}
	\begin{split}
	c_d^2&{}=\frac{n}{2m}\frac{1}{\kappa_d}\\
	c_s^2=\frac{n}{2m}\bigg[&{}1-4\frac{n_{drag}}{n}\bigg]\frac{1}{\chi_s}\; .
	\end{split}
	\label{ABv}
	\end{equation}
	Results (\ref{ABHDn}-\ref{ABHDs}) are well suited to explore the spin drag effect in the fast regime characterized by times shorter than the inverse of the frequencies $c_n q$ and $c_s q$ of the two sound  modes. By taking the difference between the two equations one finds that  the dynamic flow  of the  second component, caused by a fast perturbation applied to the first one,  takes the form  
	\begin{equation}
	\frac{d\langle F(2)\rangle}{dt}^{fast}=\lambda t\frac{ q^2}{m}\frac{n_{drag}}{n}
	\label{ABf}
	\end{equation}
	revealing explicitly  that the spin drag effect is  proportional to the drag density $n_{drag}$ entering the equation of continuity (\ref{ABHD3}) for the spin current. The dynamic flow of the first component instead follows the  law
	\begin{equation}
	\frac{d\langle F(1)\rangle}{d t}^{fast}=\lambda t\frac{ q^2}{m}\bigg[\frac{1}{2}-\frac{n_{drag}}{n}\bigg] \; .
	\label{ABfn}
	\end{equation}
\begin{table}[t]
		\begin{center}
			\caption{Mixtures without coherent coupling. Momentum-$q$ dependence of strength, excitation energy and contribution to the f-sum rule $m_1$, for excitation operators of the form $F(q)=\sum_k\cos(q x_k)$. The q-dependence is given for low and high frequency excitations in both the density and spin channels. Results hold for Bose mixtures exhibiting the Andreev-Bashkin effect and normal  Fermi liquids.\\[1ex]}
			\label{twithoutcc}
			\begin{tabular}{c c c c c c c c}
				\hline
				\\[-1ex]
				& \multicolumn{2}{c}{{\bf density}}& & & &\multicolumn{2}{c}{{\bf spin}}\\
				& \multicolumn{1}{c}{low} & \multicolumn{1}{c}{high} & & & & \multicolumn{1}{c}{low} & \multicolumn{1}{c}{high}\\[+1ex]
				\cline{2-3}\cline{7-8}
				\\[-1ex]
				\boldmath$|\langle 0|F(q)|n\rangle|^2$\unboldmath & $q$ & $q^4$ & & & & $q$ & $q^2$ \\[+1ex]
				\boldmath$\omega_{n0}$\unboldmath & $q$ & const & & & & $q$ & const \\[+1ex]
				\boldmath$m_1$\unboldmath& $q^2$ & $q^4$ & & & & $q^2$ & $q^2$\\[+1ex]
				\hline
			\end{tabular}  
		\end{center}
	\end{table}	
	In 3D the value of the drag density $n_{drag}$ in a weakly interacting Bose mixture was calculated in \cite{Fil} by developing the beyond mean field  Lee-Huang-Yang formalism in the presence of two moving Bose-Einstein condensates. In the case of gases interacting with equal intra-species scattering lengths $a_{11}=a_{22}\equiv a$ the authors of \cite{Fil} found the result
	\begin{equation}
	n_{drag}= \frac{n}{2\sqrt{2}}\frac{a_{12}^2}{a^2}\sqrt{n a^3}
	\label{Fil}
	\end{equation}
	which explicitly reveals that the spin drag effect is caused by the inter-species scattering length $a_{12}$ and is fixed by the value of the gas parameter $na^3$. The effect is very tiny  in the available 3D quantum mixtures. Larger effects are predicted to occur  in one dimensional configurations  \cite{Giorgini} or in the presence of an optical lattice \cite{lattice1,lattice2,contessi2020collisionless}.
	
	If instead of the periodic perturbation one applies the dipole perturbation $x$ to the first component, the velocities $V_1$ and $V_2$ acquired by the center of mass of two components satisfy the useful equation
	\begin{equation}
\bigg( \frac{V_2}{V_1}\bigg)^{fast}= \frac{\langle n_{drag}/n\rangle}{\langle 1/2-n_{drag}/n\rangle} 
\label{FAB}
\end{equation}
depending explicitly on the drag density, the averages taking into account the possible inhomogeneity of the equilibrium density profile.

\section {Interacting Fermi mixtures in the normal phase}
	
	The collisionless fast spin drag effect  discussed in this paper is not a phenomenon restricted to bosonic mixtures, nor to superfluids. Indeed, the relevant time scale of the phenomena considered in this work is very short compared to one characterizing the low frequency dynamics, where superfluid and quantum statistical effects are important. As we will show in this section, fast spin drag is also exhibited by interacting Fermi mixtures in the normal (non superfluid) phase \cite{amico2020roadmap}. For simplicity, here, we will discuss the case of spin-1/2  interacting mixtures. Similarly to the case of the Andreev-Bashkin effect, the phenomenon is directly related to the fact that the low energy excitations of the system, as described by Landau theory of normal Fermi liquids, do not exhaust the energy weighted f-sum rule   in the spin channel.  
	
	Landau's theory is a semi-phenomenological theory which describes the macroscopic behavior of fermionic mixtures  at low temperatures. The theory is based on the proper kinetic equation for the quasi-particle distribution function (see, for example, \cite{Pines}), which allows for the  investigation of  the proper elementary excitations of the system in the collisionless regime (collective zero-sound as well as particle-hole excitations).  The same kinetic equation allows for the calculation of the dynamic response function in both the density and spin channel and is consequently well suited to investigate the response of the system to  fast perturbations of special  interest for the present work.  The theory accounts for in-phase and out-of-phase interaction effects through the proper inclusion of the so-called Landau parameters $F_\ell$ and $G_\ell$, respectively, where $\ell$ indicates the angular nature of the deformation of the Fermi surface in momentum space.

The results for the dynamic response function predicted by Landau theory (LT) allow for the determination of the energy weighted sum rule in the density as well as in the spin channel. These can be directly derived starting from the equation of continuity in the density and spin density channels
\begin{equation}
\partial_t \delta n + {\bf \nabla}\cdot{\bf j}_d^0=0
\label{ECDF}
\end{equation}
 and 
\begin{equation}
\partial_t \delta s + {\bf \nabla} \cdot\bigg(\frac{1+G_1/3}{1+F_1/3}{\bf j}_s^0\bigg)=0\; .
\label{ECSF}
\end{equation}
In the above equation we have defined the bare current density ${\bf j}_d^0({\bf r},t)$ and spin current density ${\bf j}_s^0({\bf r},t)$ as the average value of the operators $\sum_k(1/2m)({\bf p}_k\delta({\bf r}_k-{\bf r})+\text{h.c.})$ and $\sum_k(1/2m)\sigma_z({\bf p}_k\delta({\bf r}_k-{\bf r})+\text{h.c.})$, respectively. Equation (\ref{ECSF}) explicitly shows that the physical spin density current ensuring the conservation for the integrated spin density is renormalized with respect to the bare value by the Landau parameters $F_1$ and $G_1$  \cite{Pines}. By applying an external perturbation of the form $\lambda f({\bf r})\exp{(-i\omega t)}$ and  $\lambda f({\bf r})\sigma_z \exp{(-i\omega t)}$ and noticing that, in the large frequency limit, the variation of the bare current is determined by the external force according to 
$m\partial_t {\bf j}^0_{n(s)}({\bf r},t)_{\omega \to \infty}=- \lambda n{\bf\nabla} f({\bf r}) \exp{(-i\omega t})$,  it is straightforward to  calculate the energy weighted sum rule, using the formalism of linear response theory \cite{Book}.  In the density channel we obtain the exact f-sum rule  result
\begin{equation}
	\begin{split}
	m_1^{LT}(F_n) = 
	\frac{1}{2 m}\int d{\bf r} |\nabla f({\bf r})|^2n({\bf r}) \; ,
	\label{m1Fdensity}
	\end{split}
	\end{equation}	
 reflecting the Galilean invariance of the theory, while in the spin channel one instead finds \cite{PhysRevLett.37.845,PhysRevLett.63.532}
\begin{equation}
	\begin{split}
	m_1^{LT}(F_\sigma) = 
	\frac{1}{2 m}\int d{\bf r} |\nabla f({\bf r})|^2\frac{1+G_1/3}{1+F_1/3}n({\bf r})
	\label{m1Fsigma}
	\end{split}
	\end{equation}
resulting in the violation of the usual  f-sum rule.  In accordance to the general considerations discussed in the previous sections, we associate the violation of the $f$-sum rule  to the fact that high energy excitations of multi-pair type, not accounted for by  Landau theory, are responsible for the remaining contribution to the energy weighted sum rule \cite{PhysRevLett.37.845,PhysRevLett.63.532}. The effect is well elucidated in Table \ref{twithoutcc} where we report the q-dependence of the low and high frequency contributions to the energy weighted sum rule in the density and spin channels in the case of the periodic perturbation $f({\bf r})=\cos (q x)$. 
In strongly interacting Fermi liquids, like liquid $^3$He, the  Landau parameters   $F_0$, $G_0$, $F_1$ and $G_1$ can be determined in a phenomenological way, by proper comparison with available experimental results for the compressibility, the magnetic susceptibility,   the heat capacity and the transverse spin diffusivity \cite{Baymbook}. Recent experiments on the Leggett-Rice effect have provided first useful information of the spin current interaction parameter $G_1$ \cite{trotzky}.  In a weakly interacting Fermi gas the Landau parameters can be calculated using perturbation theory. For the relevant  $\ell=1$  parameters entering the expression for the energy weighted sum rule (\ref{m1Fsigma}) one finds the result (see, for example, \cite{PRLrecstr2011}) 
	\begin{equation}
	\begin{split}
	F_1=\frac{8}{5\pi^2}(7\ln2-1)(k_F a)^2\\
	G_1=-\frac{8}{5\pi^2}(2+\ln2)(k_F a)^2
	\end{split}
	\end{equation}
	where $a$ is the s-wave scattering length and $k_F$ is the Fermi wave vector (for a 3D system $k_F=(3\pi^2 n)^{1/3}$). Both parameters depend on $k_F a$ quadratically, revealing explicitly that  the  spin drag effect has a typical beyond mean field nature, similarly to the case of the AB effect discussed in the previous section. 
	
	Taking the difference (\ref{FnFs}) between the density and spin density response functions and results (\ref{m1Fdensity}) and (\ref{m1Fsigma}) for the corresponding energy weighted sum rules in the case of Fermi mixtures, in the case of uniform Fermi mixtures one finds that, by  applying a perturbation of the form $-\lambda \cos(q x)\Theta(t)$ to the component $1$  of the mixture, in the fast regime the component $2$, which is  not  directly affected by the perturbation, is put  in motion according to
	\begin{equation}
		\frac{d\langle F(2)\rangle}{dt}^{fast}=\frac{\lambda t}{4}\frac{ q^2}{m}\frac{(F_1-G_1)/3}{1+F_1/3}
		\label{Ff}
		\end{equation} 
as a consequence of the  spin drag effect, which   depends explicitly on the two interaction parameters $F_1$ and $G_1$ and vanishes only if $F_1=G_1$. The time dependence of $\delta\langle F(1)\rangle$ is instead given by 
\begin{equation}
		\frac{d\langle F(1)\rangle}{dt}^{fast}=\frac{\lambda t}{4}\frac{ q^2}{m}\frac{2+(F_1+G_1)/3}{1+F_1/3}
		\label{Ff1}
			\end{equation}
		
In the case of a dipole perturbation one finds that the ratio between the  velocities $V_1$ and $V_2$ acquired by the center of mass of the two components after the fast pulse applied to the component 1 takes the same form
\begin{equation}
\bigg( \frac{V_2}{V_1}\bigg)^{fast}= \frac{\langle n_{drag}/n\rangle}{\langle 1/2-n_{drag}/n\rangle} 
\label{FLT}
\end{equation}
as for the Andreev-Bashkin effect with
\begin{equation}
n_{drag}= \frac{F_1-G_1}{3+F_1}\frac{n}{4} \;.
\label{ndragFL}
\end{equation}
The Landau parameters $F_1$ and $G_1$ (and consequently the drag density) should satisfy two important conditions. A first condition, corresponding to the stability criterion against the spontaneous formation of spin currents, is $G_1>-3$, i.e. $n_{drag}<n/4$. This corresponds to ensuring that the   f-sum rule (\ref{m1Fsigma}), calculated within Landau theory, be positive. A second condition is  that the Landau f-sum rule (\ref{m1Fsigma}) does not exceed the microscopic value $(1/2 m)\int d{\bf r} |\nabla f({\bf r})|^2n({\bf r})$. This imposes the condition $G_1<F_1$ \cite{BP}, i.e. $n_{drag}>0$ and is equivalent to requiring that the multi-pair contribution to the full f-sum rule be positive.

Our results can be used to provide quantitative estimates for the spin drag effect in a dilute Fermi gas interacting with a negative scattering length. For example, choosing  $k_Fa=-0.5$, one predicts an effect of about $4$\% for the ratio $V_2/V_1$. At the same time the critical temperature $T_c$ for the transition to the BCS superfluid phase is still very small compared to the Fermi temperature and consequently Landau theory is reasonably applicable. When the value of $|a|$ increases, approaching the unitary regime, the value of $T_c$ becomes comparable to the Fermi temperature and experiments on spin drag could be consequently employed to test the limits of applicability  of Landau's theory of Fermi liquids.

\section{Polaron}

In this last section we apply the formalism of fast spin drag to the problem of the polaron, where the drag, as expected, is determined by the value of the  effective mass $m^*_P$ which characterizes the kinetic energy term  $p^2/2m^*_P$ of the polaron Hamiltonian. 
The problem of the polaron, investigated a long time ago in the case of dilute mixtures of $^3$He in liquid $^4$He \cite{BP},  has  recently attracted large experimental and theoretical interest in the community of ultracold quantum gases (see, for example, \cite{Zwierlein2020} and references therein). 

A convenient way to explore the problem starts from   the equations of continuity for the densities $n_1$ and $n_2$ of two components of the mixture. These  can be derived employing a Galilean invariant expression for the energy functional including a velocity dependent interaction  term proportional to the square   $({\bf v}_1({\bf r})-{\bf v}_2({\bf r}))^2$  of the difference between the velocity fields of the two components. The equations of continuity then take the form
\begin{equation}
\partial_t n_1+ {\bf \nabla}\cdot\bigg(\frac{m_1}{m_1^*}n_1{\bf v}_1+ \frac{m_2}{m_1}(1-\frac{m_2}{m_2^*})n_2{\bf v}_2\bigg)=0
\label{n1Pol}
\end{equation}
and 
\begin{equation}
\partial_t n_2+{\bf \nabla} \cdot \bigg(\frac{m_2}{m_2^*}n_2{\bf v}_2+ \frac{m_1}{m_2}(1-\frac{m_1}{m_1^*})n_1{\bf v}_1\bigg)=0 \; ,
\label{n2Pol}
\end{equation}
where $m_1^*$ and $m_2^*$ are   the effective masses characterizing the renormalization of the bare masses caused by the presence of interaction between the two mixture components.
In the case of superfluid gases, these velocity fields  take the form ${\bf v}_1=n_1\nabla\phi_1/m_1$ and ${\bf v}_2=n_2\nabla\phi_2/m_2$, respectively, with $\phi$ the usual phase of the order parameter.   The effective masses entering Eqs.(\ref{n1Pol}-\ref{n2Pol}) satisfy  the  relationship 
\begin{equation}
n_1m_1\bigg(1-\frac{m_1}{m_1^*}\bigg) = n_2 m_2\bigg(1-\frac{m_2}{m_2^*}\bigg)
\label{relm1m2}
\end{equation}
following from the Galilean invariant nature of the interaction term \cite{recati}. It is actually easy to check that the total mass density $m_1n_1+m_2n_2$ satisfies the equation of continuity 
\begin{equation}
\partial_t (m_1n_1+m_2n_2) + {\bf \nabla}\cdot (m_1n_1{\bf v}_1 + m_2n_2{\bf v}_2)=0 \; ,
\label{Masscontinuity}
\end{equation}
independent of interaction effects. 
The above formalism applies to Bose as well as to Fermi superfluids (in the latter case the bare mass corresponds to twice the atomic value). 

A fast kick caused by the perturbation $\lambda f({\bf r})$ applied only  to the majority component (hereafter called component $1$) produces a non vanishing value of its velocity field, given by ${\bf v}_1=\lambda t \nabla f/m_1$, while  ${\bf v}_2=0$ so that, by evaluating the effect of the perturbation on the average value of the quantity $f({\bf r})$, relative to the  components $1$ and $2$, one finds:
\begin{equation}
\frac{d \langle F(1)\rangle}{dt}^{fast} =  \frac{\lambda t}{m_1N_1} \int d{\bf r} |\nabla f({\bf r})|^2 n_1({\bf r}) \frac{m_1}{m^*_1}
\label{P1}
\end{equation}
and 
\begin{equation}
\frac{d \langle F(2)\rangle}{dt}^{fast} =  \frac{\lambda t}{m_1N_2} \int d{\bf r} |\nabla f({\bf r})|^2 n_2({\bf r}) \bigg(1-\frac{m_2}{m^*_2}\bigg) \; .
\label{P2}
\end{equation}

In the case of a dipole perturbation ($f({\bf r}) =x)$ one finds that  the ratio   between the velocities $V_2=d\langle x_2\rangle/dt$ and   $V_1= d\langle x_1\rangle/dt$ acquired by the center of mass of the minority and majority components,    takes the   form
\begin{equation} 
\frac{V_2}{V_1}= \frac{\langle 1-\frac{m_2}{m_2^*}\rangle}{\langle \frac{m_1}{m_1^*}\rangle} \simeq  \langle 1-\frac{m_2}{m_2^*}\rangle
 \label{P1}
\end{equation}
where the average, as usual, takes into account the possible inhomogeneity of the medium and, in the last equality, we have set $m_1^*\simeq m_1$ which follows from Eq. (\ref{relm1m2}) in the limit $n_2 \ll n_1$ corresponding to the case of few impurities.
 
The above equations provide the sought result for the drag effect in terms of the polaron effective mass $m_2^*\equiv m_P^*$, which might hopefully become an efficient tool for its measurement.

\section{Conclusions}

In this paper we have discussed the peculiar collisionless spin drag effect associated with a fast perturbation applied to one of the components of the mixture and resulting in the motion of the second one. We have shown that the  spin drag effect is directly related to the behavior of the energy weighted  sum rule in the density and spin density channels and in particular to the distinction between the high and low energy excitations contributing to the above sum rules. Novel results are derived in the case of coherently coupled Bose-Einstein condensates, superfluid mixtures exhibiting the Andreev-Bashkin effect, normal Fermi fluids and the polaron problem. While in the first case spin drag is directly associated to the presence of the coherent coupling which modifies   the f-sum rule in the spin channel, in the other cases fast spin drag  is the consequence of beyond mean field effects.   We hope that this work will stimulate novel experimental investigations, providing further insight on the transport phenomena exhibited by quantum mixtures.  Natural extensions of our approach include quantum mixtures in the presence of  periodic optical potentials.

Stimulating discussions with Stefano Giorgini and Alessio Recati are acknowledged. This project has received funding from the European Union's Horizon 2020 research and innovation programme under grant
agreement No. 641122 "QUIC", from Provincia Autonoma di Trento,  and the FIS$\hbar$
project of the Istituto Nazionale di Fisica Nucleare. F. C. acknowledges the support from the 80'PRIME-international CNRS programme.
	
	\bibliography{references}
\end{document}